\documentclass{ptapap}

\newcommand{\hvezda}{HD\,37776}

\author{Zden\v{e}k Mikul\'a\v{s}ek}[UTFA]
\author{Ji\v{r}\'{i} Krti\v{c}ka}[UTFA]
\author{Matthew E. Shultz}[UPA]
\author{Gregory W. Henry}[CEIS]
\author{Milan Prv\'ak}[UTFA]
\author{Alexandre David-Uraz}[UPA]
\author{Jan Jan\'{i}k}[UTFA]
\author{Miloslav Zejda}[UTFA]
\author{Iosif I. Romanyuk,}[SAO]
\author{MOBSTER collaboration}[]
\affil[UTFA]{Department of Theoretical Physics and Astrophysics, Masaryk University,  Kotl\'a\v{r}sk\'a 2, 61~387 Brno, Czech Republic}
\affil[UPA]{Department of Physics and Astronomy, University of Delaware, 217 Sharp Lab, Newark, DE 19716, USA}
\affil[CEIS]{Center of Excellence in Information Systems, Tennessee
State University, Nashville, Tennessee, USA}
\affil[SAO]{Special Astrophysical Observatory of the RAS, Nizhnii Arkhyz, Karachai-Cherkessian Republic, 369167, Russia}
\title{What’s New with Landstreet’s Star HD 37776 (V901 Ori)?}

\begin{document}

\maketitle

\begin{abstract}

HD\,37776 (V901\,Ori, B2\,Vp), also known as Landstreet's Star, is possibly the most remarkable magnetic chemically peculiar (mCP) star known. Zeeman Dop-pler Imaging revealed this young, rapidly rotating star's surface magnetic field to be not only the strongest ($\sim 30$ kG) of the He-strong class of hot mCP stars but also by far the most topologically complex. In contrast to the overwhelming majority of mCP stars, which are well described by tilted dipoles, Landstreet's Star's non-axisymmetric surface magnetic field is entirely dominated by high-order spherical harmonics. It is one of the handful of stars for which rotational period change has been measured, and over the past two decades of monitoring, the object has demonstrated an unexpected acceleration in its rotation that so far defies explanation. Recently acquired TESS data have provided a photometric data set of unprecedented precision. These data have revealed a highly stable yet multi-featured light curve, making Landstreet's Star the prototype of hot mCP stars whose light curves are difficult to reproduce using the standard model of chemical/photometric spots modulated by solid-body rotation.

\end{abstract}

\section{Introduction}

\hvezda\ (V901 Ori), known as Landstreet’s Star, is a young hot star (B2 V) embedded in the emission nebula IC 432. The star belongs to the OB1 Orion association \citep{land07}. This He-strong chemically peculiar star \citep{nissen76} shows rotationally modulated variations with a period of 1.5387\,d in the equivalent widths of He~{\sc i} and Si~{\sc ii} lines \citep{pedersen77,walborn82}, and in light \citep{adelman97}. A magnetic field displaying rotationally modulated changes was revealed by \citet{borra79} and confirmed by \citet{thompson85}.

\citet{chochlova00} then derived the magnetic field geometry and maps of chemical elements distribution. \citet{krta07} found that the light variations can be explained by energy redistribution from ultraviolet regions due to line/band blanketing caused mainly by silicon. Bright spots coincide with over- and under-abundances of Si and He respectively.

\begin{figure}
\centering\includegraphics[width=0.7\textwidth]{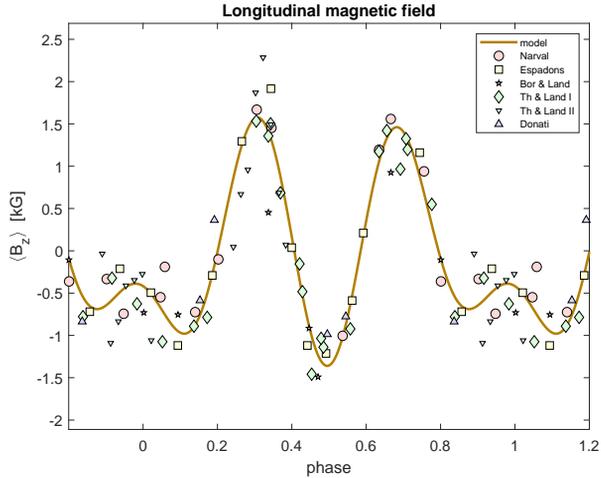}
\caption{\hvezda: phase curve of the mean longitudinal magnetic field $\langle B_z \rangle$ is triple-wave. The origins of the spectropolarimetric measurements are described in the legend. Espadons and Narval are from \cite{2018MNRAS.475.5144S}. The remaining measurements, assigned as triangles, diamonds, and stars, were taken from \citet{donati97}, \citet{thompson85}, and \citet{borra79}. Symbol size corresponds to weights.}
\label{fig:1}
\end{figure}

\citet{borra79} ascertained that the $\langle B_z \rangle$ phase curve of \hvezda\ is double-wave, which was confirmed by \citet{thompson85}. The fact was explained by unusual magnetic field geometry by \citet{bohlender94}, then \citet{kocuchov11} which
claimed: `the surface magnetic field is not only the strongest (about 30\,kG) of all non-degenerate stars but also by far the most topologically complex. Its surface magnetic field is dominated by high-order multipoles'. All $\langle B_z \rangle$ measurements available to date are presented in Fig.\,\ref{fig:1}.

\section{Period evolution}

Based on an analysis of photometric data obtained by 2005, \citet{mik05} announced that the rotation of \hvezda\ was likely to be decelerating. With three years of additional data, \citet{mik08} published an in-depth periodic study of \hvezda\ using all available photometric, spectroscopic, and spectropolarimetric observations \citep[a detailed description of their method can be found in][]{mik16}. Their analysis revealed that the deceleration of the rotational period was decreasing, such that it should change to acceleration in the near future. This was confirmed not only for \hvezda\ but also for the very fast-rotating mCP star CU\,Vir \citep{mik11,mik16}.

\begin{figure}
\centering\includegraphics[width=0.7\textwidth]{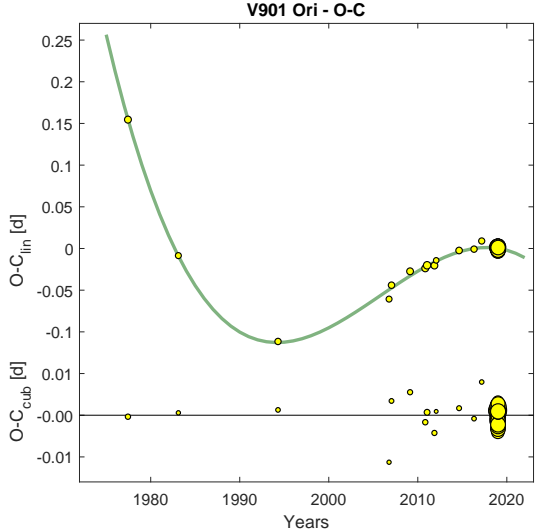}
\caption{\hvezda: The new O-C diagram represented by yellow circles clearly shows that the dependence can be described very well by a cubic polynomial. We do not yet know the cause(s) of the period change.}
\label{fig:2}
\end{figure}

The application of our method of monitoring the rotation period changes using all available data containing phase information, including 14\,871 TESS observations, confirmed that star has been accelerating since the year $2005.8\pm0.4$. The data cover the interval 1975--2018. We used 19\,116 photometric observations in the range 350--790 nm in 6 filters, 461 equivalent widths of 6 He\,I lines between 444 and 706 nm, and 75 observations of the longitudinal magnetic field from 5 sources. Strict tests show that the appearance of the phase curves has not changed in the last 50 years \citep{mik18}. Since 2006, the period has shortened by 5.5 s, to today's value of $P_0 = 1.5387394(24)$\,d (see Fig.\,\ref{fig:2}).

All types of mCP stars are represented among stars with similar behavior. The latest addition is the rapidly accelerating He-weak mCP star HD 142990 \citep{shultz19a}. The cause(s) of this remain unclear; the mechanism of torsional oscillations suggested by \citet{krta17} cannot be generalized to all of the known cases.

The expectable cyclicity of period changes or oscillations of the instant period around some mean value is proven so far only for CU Vir = HD\,124224 \citep[the cycle length $\sim 66$\,years, $\frac{\mathrm d^3\!P}{\mathrm dt^3}=3.25\times10^{-16}$\,d$^{-2}$;][]{mik17,krta19}.
For \hvezda\ $\frac{\mathrm d^3\!P}{\mathrm dt^3}$ is zero, so we only estimate that the cycle length is longer than 120 years (see Fig.\,\ref{fig:2}).
\begin{figure}
\centering\includegraphics[width=0.8\textwidth]{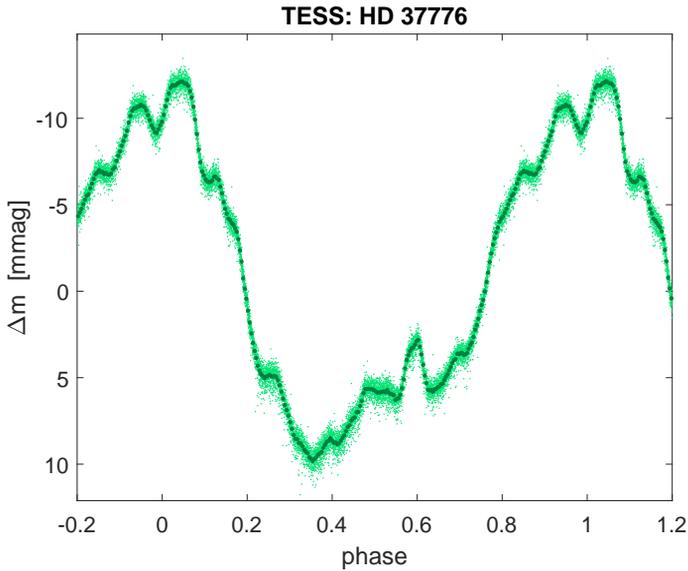}
\caption{\hvezda: The phased TESS light curve of HD\,37776 shows a number of details that have been faithfully repeated throughout the whole TESS observation. The light curve is represented by normal points (dark green).}
\label{fig:3}
\end{figure}

\section{Light curve distortions and their nature}
\label{warps}

\begin{figure}
\centering\includegraphics[width=0.99\textwidth]{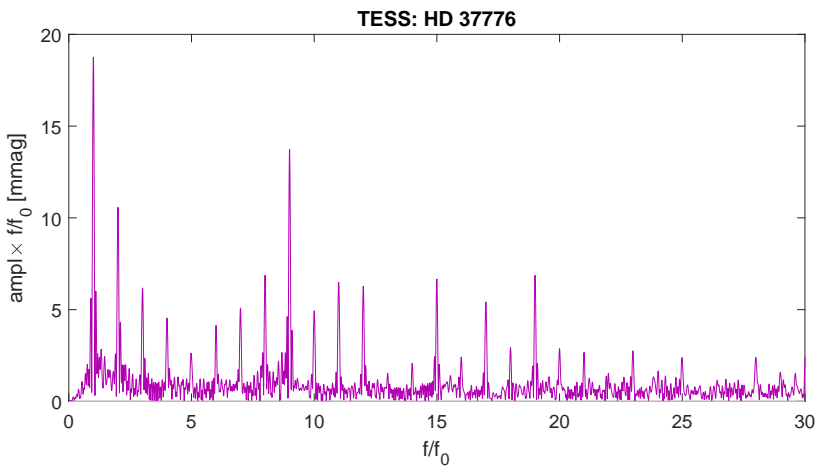}
\caption{The newly developed warp periodogram, briefly described in Sect.\,\ref{warps}, has proven to be very useful for detection of details in the light curve of \hvezda.}
\label{fig:4}
\end{figure}

The high-precision TESS photometry ($\sigma = 0.43$ mmag) revealed unexpected details in the light curve of \hvezda, which until now has been considered to be a smooth, single-wave curve - see Fig.\,\ref{fig:3}. To prove the presence and quantify the warps in the light curves of periodically variable stars, we introduce a unique tool: the warp-periodogram, shown for the case of HD\,37776 in Fig.\,\ref{fig:4}. The frequency is plotted on the $x$-axis in units of the rotation frequency $f_0$. The $y$-axis plots the product of the effective amplitude and the modified frequency ($f/f_0$), the quantity proportional to the amplitude of the time derivative changes.

\begin{figure}
\centering\includegraphics[width=0.8\textwidth]{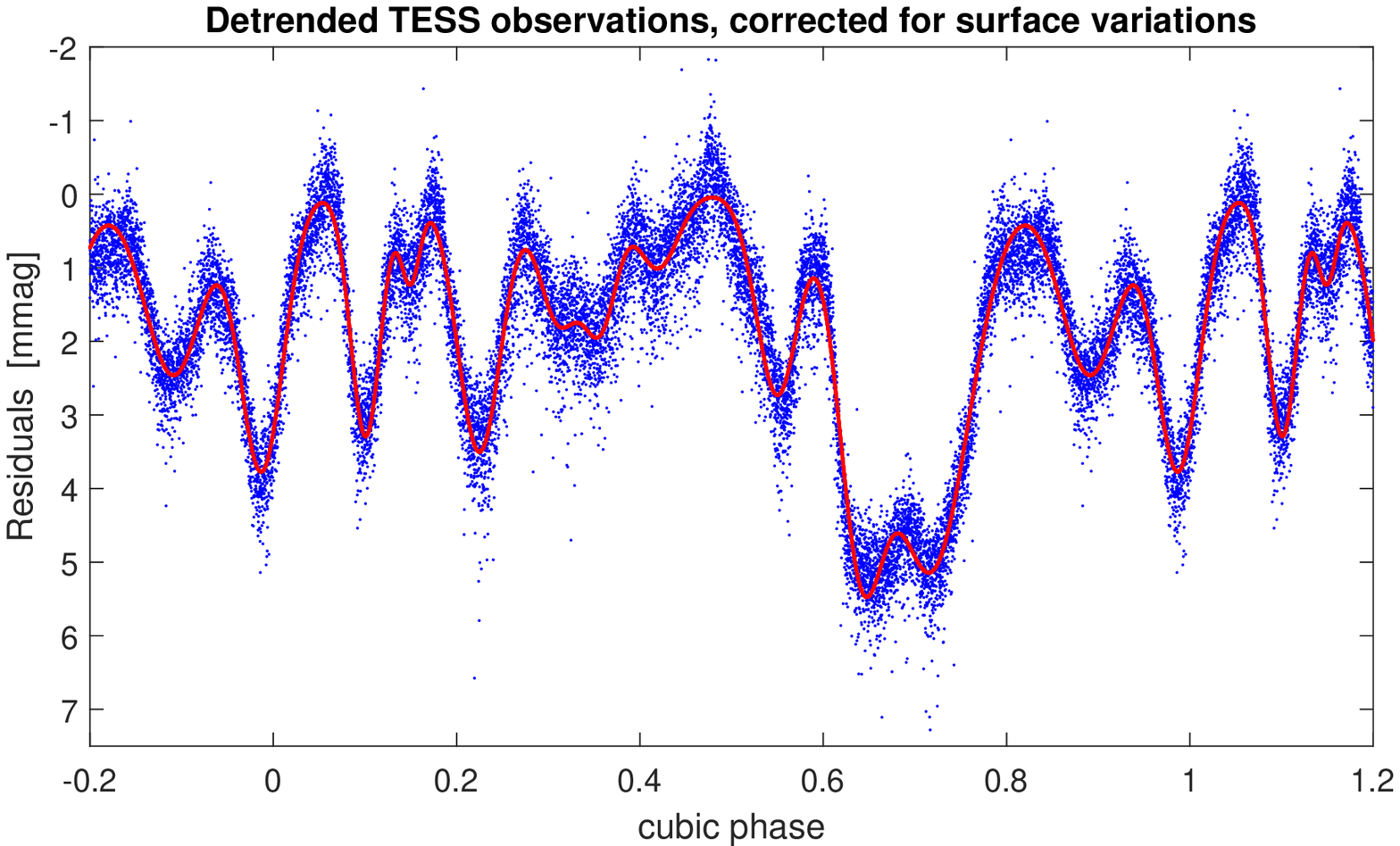}
\caption{\hvezda: Phenomenological modeling of warps in the light curve assuming that they are caused by drops in brightness due to transits of structures in the co-rotating stellar magnetosphere.}
\label{fig:5}
\end{figure}

It can be seen from the warp-periodogram of HD\,37776 that peaks up to the thirtieth harmonics are detectable in the TESS data, confirming the incredibly complex structure of the star's light curve. Structures from the fourth to the eighteenth overtones are dominant. Similar cases, albeit not so dramatic, have been found in light curves obtained with ultra-precision photometry in several tens of predominantly hot mCP stars.

\begin{figure}
\centering\includegraphics[width=0.75\textwidth]{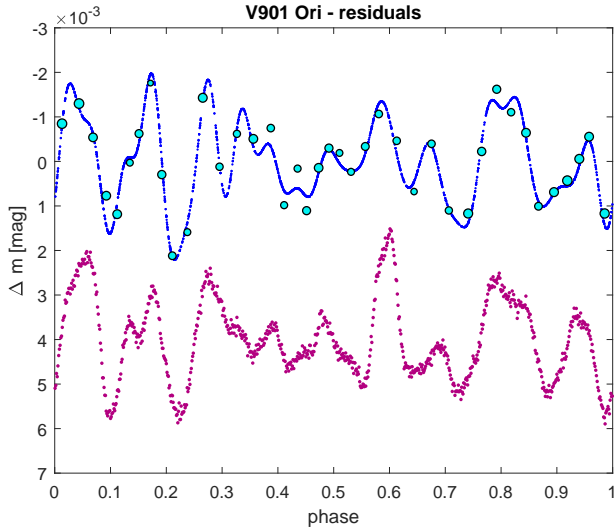}
\caption{\hvezda: The residuals in the light curve of the star derived from all available photometry before the TESS observation represented by the normal points (circles) and the light curve model (blue points -- the top of the graph) show a similar pattern to the residuals in the TESS observation, represented also by normal points (magenta). We conclude that warps are more or less persistent features in the phase light curve.}
\label{fig:6}
\end{figure}

Careful analysis of the \hvezda\ photometry from the past 50 years shows the light curve warps observed in the TESS light curve are more or less persistent. The amplitudes of warps in different filters correspond well with amplitudes of the warps in the TESS data (Fig.\,\ref{fig:6}). Our attempts to explain warps using a rotating photospheric model with photometric spots were entirely unsuccessful. Not even using a model with many small but highly contrasting spots is able to reproduce the complex structure of the light curve. We are convinced that we need to look for the cause of the low-amplitude, high-harmonic variability within the co-rotating circumstellar environment -- i.e.\ within the stellar magnetosphere. Observations of variable H$\alpha$ emissions in some Bp stars prove that their magnetospheres are filled with inhomogeneously distributed ionized gas.

We can speculate that in a complex co-rotating magnetosphere such as that expected for HD\,37776, there may be traps filled with opaque plasma. Light variations can occur when the clouds eclipse the stellar disk \citep{landstreet78,townsend05}, due to scattering from the clouds when they are projected beside the disk, or by occultation of the clouds by the star. The change in brightness would then depend on the distance of the clouds from the axis of rotation, as well as on their projected shape and size. The light curves can be phenomenologically modeled.

\section{Other warped mCP stars and the enigmatic $\sigma$ Ori E}

\begin{figure}
\centering\includegraphics[width=0.75\textwidth]{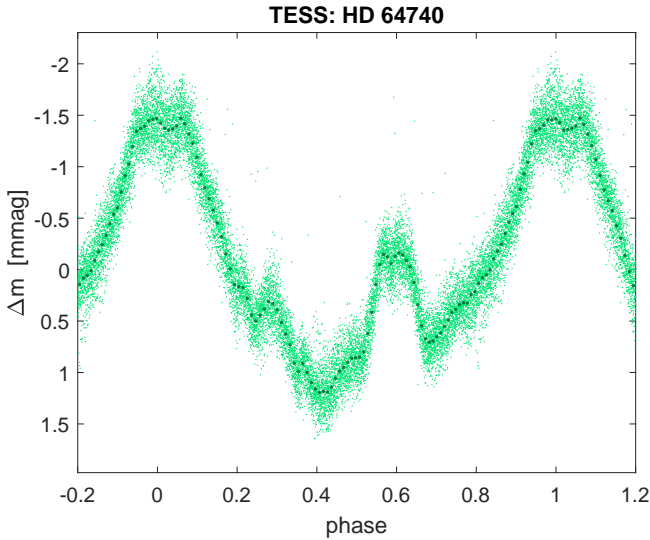}
\caption{TESS light curve of the Helium-strong B2\,V magnetic CP star HD\,64740 \citep{shultz19b}. Similarly to \hvezda\, this star displays warps in its TESS light curve.}
\label{fig:7}
\end{figure}

It seems that warping in the \hvezda\ light curve is not a unique phenomenon, but is commonly found in the light curves of other magnetic CP stars, many of which also display H$\alpha$ emissions. It is true e. g.\ in the case of HD\,64740 \citep[][and citations therein]{shultz19b}, see Fig.\,\ref{fig:7}. At present, we have recorded several dozen warped mCP stars, and expect their number to increase with the increasing number of stars observed by TESS.

Thus, the study of warps promises to become an independent diagnostic for the magnetospheres of hot magnetic stars. The method opens the possibility of doing exciting magnetospheric science based on statistically significant samples of magnetic stars.

\begin{figure}
\centering\includegraphics[width=0.65\textwidth]{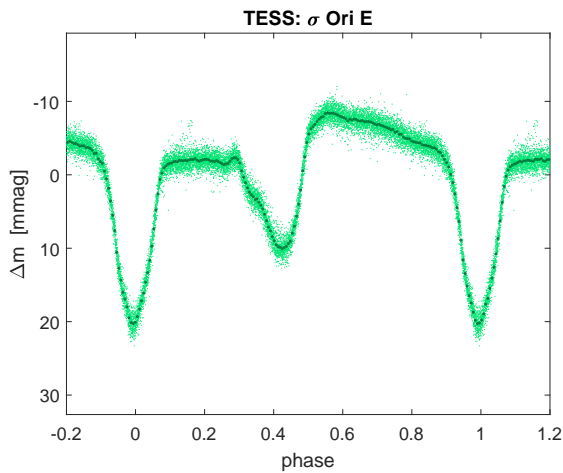}
\caption{$\sigma$ Ori E = HD\,37429: The phased TESS light curve shows two declines at phases 0 and 0.4 due to eclipsing by the star's corotating magnetosphere \citep{townsend10,oksala15}. The variation between phases 0.6 and 0.9 cannot be reproduced by a photospheric model, and may also be circumstellar in origin.}
\label{fig:8}
\end{figure}

Our last remark: $\sigma$ Ori E is a notorious hybrid of a Be star with strong H$\alpha$ emission and a magnetic He-rich star with a strong magnetic field \citep[discovered by][]{landstreet78} and a large magnetosphere. The star's light curve is well known to show evidence of periodic eclipses of the star by its magnetosphere \citep[e.g.][]{townsend10,oksala15,mik18}, and clearly qualifies as an extremely warped light curve (see Fig.\,\ref{fig:8}). \cite{oksala15} noted that while the eclipses are well-reproduced by the Rigidly Rotating Magnetosphere model developed by \cite{townsend05}, the out-of-eclipse variability could not be reproduced by a careful photometric model of photospheric brightness variations due to chemical abundance spots mapped via Doppler Imaging. They speculated that scattering of light from the magnetospheric plasma when not eclipsing the star may be the explanation for the otherwise inexplicable photometric variability. If so, this suggests that circumstellar plasma may also result in warping in other stars.

%The biggest challenge is the nature of an atypical (yet strictly periodic) optical light curve, which cannot be explained by the standard oblique rotator model. We should these eclipses also qualify as gigantic warps .
%\citet{townsend10,oksala15,mik18}, and others: the result of the presence of rigidly rotating magnetosphere driving nearby circumstellar material.

\section{Conclusions}
Landstreet’s Star = HD 37776 = V901 Ori is an object revealing never-ending surprises. New research confirmed the record-breaking strength and complexity of the global magnetic field, the presence of a co-rotating magnetosphere filled in with ionized gas, and the present acceleration of its rotation, the fastest amongst all mCP stars.

We revealed the incidence of a dozen well-defined light curve warps with typical half-widths of 0.054 in-phase and depth of 2.7 mmag, persistent during the last 44 years. The warps in the light curve of \hvezda\  cannot be explained in the frame of a purely photospheric oblique rotator model with surface abundance spots. Instead, they might be the result of scattering from and eclipsing of the star by opaque structures in the co-rotating magnetosphere. The LC warping of Landstreet's Star is the most striking example of a more general phenomenon. Analysis of light curve warps promises to be a new tool for magnetospheres diagnostics.

\acknowledgements{Z. Mikul\'a\v{s}ek and J. Jan\'{i}k were granted by GA\v{C}R 18-05665S. M. E. Shultz acknowledges financial support from the Annie Jump Cannon Fellowship, supported by the University of Delaware and endowed by the Mount Cuba Astronomical Observatory.}

\bibliographystyle{ptapap}
\bibliography{pasmik3}

\end{document}